# Teleparallel Conformal Killing Vector Fields of LRS Bianchi Type V Spacetimes in Teleparallel Gravity

Suhail Khan, Tahir Hussain[*] and Gulzar Ali Khan

Department of Mathematics, University of Peshawar, Khyber Pakhtoonkhwa, Pakistan

[*]Email: lec.tahir@yahoo.com

## Abstract

The aim of this paper is to explore teleparallel conformal Killing vector fields (CKVFs) of locally rotationally symmetric (LRS) Bianchi type V spacetimes in the context of teleparallel gravity and compare the obtained results with those of general relativity. The general solution of teleparallel conformal Killing's equations is found in terms of some unknown functions of $t$ and $x$, along with a set of integrability conditions. The integrability conditions are solved in some particular cases to get the final form of teleparallel CKVFs. It is observed that the LRS Bianchi type V spacetimes admit proper teleparallel CKVF in only one case, while in remaining cases the teleparallel CKVFs reduce to teleparallel Killing vector fields (KVFs). Moreover, it is shown that the LRS Bianchi type V spacetimes do not admit any proper teleparallel homothetic vector field (HVF).

## 1. Introduction

After the final presentation of Einstein's theory of General Relativity (GR) as a new theory for the gravitational field, several attempts were made to unify gravitation and electromagnetism. One of such attempts was made by Weyl in 1918 [1], whose proposal did not succeed but it led to the gauge transformations and gauge invariance. After about ten years, another similar attempt was made by Albert Einstein [2], which was based on mathematical structure of teleparallelism. This idea introduced a field of orthonormal bases on tangent spaces at each point of a four dimensional spacetime, known as tetrad. Like Weyl's work, Einstein's attempt of unification also did not succeed but it introduced certain concepts that remained important till now.

After three decades, Moller [3] gave a new strength to the Einstein's idea, not for the purpose of unification, but in pursuing the gauge theory of gravitation. Pellegrini and Plebanski [4] followed the Moller's idea to find a Lagrangian formulation for teleparallel gravity. The gauge theory for translation groups was formulated by Hayashi and Nakano in 1967 [5]. A connection between gauge theory for translation groups and teleparallelism was presented by Hayashi [6]. Following this, Hayashi and Shirafuji [7] made an attempt to unify these developments. In this way the theory of GR was supplemented by teleparallel gravity, however both the theories are found to be equivalent and curvature and torsion provide similar description of the gravitational interaction. In contrast to GR, which involves only curvature, the teleparallel gravity involves non-vanishing torsion and vanishing curvature. Moreover GR is based on Riemannian geometry while teleparallel gravity is based on Weitzenböck geometry [8].

There has been many contributions from different authors to GR and teleparallel gravity. In particular, the study of spacetime symmetries is a significant area of recent research in both the theories. A symmetry of a spacetime is a vector field whose associated local diffeomorphisms preserve some geometrical feature of the spacetime [9]. Spacetime symmetries are used to understand the natural relationship between geometry and matter given by the well known Einstein's field equations (EFEs). Spacetime symmetries are helpful in finding the exact solutions of EFEs and in the classifications of spacetimes. Also, they have a close link with conservation laws [10].

The symmetry of our interest in this paper is conformal symmetry, which preserves the metric of a spacetime up to a conformal factor. In GR, this type of symmetry satisfies the relation[9]:

$$\pounds_\xi g_{mn} = g_{mn,r}\xi^r + g_{mr}\xi^r_{,n} + g_{nr}\xi^r_{,m} = 2\psi\, g_{mn}, \tag{1.1}$$

where $\pounds_\xi$ represents Lie derivative operator along the vector field $\xi$ and $\psi$ is some smooth real valued function which depends on the chosen coordinate system. The commas in subscript are used to represent partial derivatives. Equation (1.1) is usually referred as conformal Killing's equation. If $\psi$ in Eq. (1.1) becomes constant, the vector field $\xi$ reduces to a HVF and if $\psi$ vanishes, the vector field $\xi$ becomes a KVF. Sharif and Amir [11] developed the teleparallel version of Lie derivative, due to which the conformal Killing's equation (1.1) took the following form.

$$\pounds^T_\xi g_{mn} = g_{mn,r}\xi^r + g_{mr}\xi^r_{,n} + g_{nr}\xi^r_{,m} + \xi^r(g_{mp}T^p{}_{nr} + g_{np}T^p{}_{mr}) = 2\psi\, g_{mn}, \tag{1.2}$$

where $£L^T_\xi$ denotes teleparallel Lie derivative operator along the vector field $\xi$ and $T^p{}_{nr}$ and $T^p{}_{mr}$ are the components of torsion tensor. The vector fields satisfying Eq. (1.2) are known as teleparallel conformal Killing vector fields.

The conformal symmetries have been widely investigated for so many spacetimes in GR [12-19] but in the framework of teleparallel gravity, these symmetries have been explored only for few spacetimes [20-23].

The Bianchi type V cosmological models are natural generalization of FRW models with negative curvature. These models are of interest because they include isotropic special cases and allow arbitrary small anisotropy at some instant of cosmic time. Because of this property, Bianchi type V models are considered as most suitable candidates for the Universe. Apart from this, Bianchi type $I, V$ and $IX$ models include the homogeneous and isotropic FRW models as their special cases according to $t = \text{constant}$, zero, negative or positive. Although homogeneous but anisotropic models are more restricted than the inhomogeneous models, they describe a number of observed phenomena quite satisfactorily [24].

In this paper, we study teleparallel CKVFs in LRS Bianchi type V spacetimes and compare our results with those obtained in general relativity [19]. The paper is organized as: In section 2, we give definitions of some basic concepts of teleparallel gravity. Teleparallel conformal Killing's equations for LRS Bianchi type V spacetimes and their general solution, along with some integrability conditions are presented in section 3. In section 4, we solve these integrability conditions for some particular values of the metric functions to get the explicit form of teleparallel CKVFs. A brief summary is given in the last section.

## 2. Basic Concepts of Teleparallel Gravity

The covariant derivative of a covariant tensor of rank two in teleparallel theory of gravitation is defined as [11]:

$$\nabla_\rho T_{\mu\nu} = T_{\mu\nu,\rho} - \Gamma^\theta{}_{\rho\nu} T_{\mu\theta} - \Gamma^\theta{}_{\mu\rho} T_{\nu\theta}, \qquad (2.1)$$

where the symbols $\Gamma^\theta{}_{\rho\nu}$ denote Weitzenböck connections, which are defined as:

$$\Gamma^\theta{}_{\mu\nu} = h_a{}^\theta \partial_\nu h^a{}_\mu. \qquad (2.2)$$

In Eq. (2.2), $h^a{}_\mu$ represents the non-trivial tetrad field. Its inverse field is denoted by $h_a{}^\mu$ and the tetrad and its inverse satisfy the following relations:

$$h^a{}_\mu h_a{}^\nu = \delta_\mu{}^\nu, \quad h^a{}_\mu h_b{}^\mu = \delta_b{}^a. \qquad (2.3)$$

The metric coefficients can be generated from the tetrad field as:

$$g_{\mu\nu} = \eta_{ab} h^a{}_\mu h^b{}_\nu, \qquad (2.4)$$

where $\eta_{ab} = \text{diag}(-1,1,1,1)$ is the Minkowski metric. The torsion tensor can be expressed in terms of Weitzenböck connections as:

$$T^\theta{}_{\mu\nu} = \Gamma^\theta{}_{\nu\mu} - \Gamma^\theta{}_{\mu\nu}. \qquad (2.5)$$

## 3. Teleparallel Conformal Killing's Equations and Their Solution

We consider LRS Bianchi type V spacetimes with the metric [25]:

$$ds^2 = -dt^2 + A^2(t)dx^2 + e^{2qx}B^2(t)[dy^2 + dz^2], \qquad (3.1)$$

where $A$ and $B$ are nowhere zero functions of $t$ only and $q \in \Re$. If $q = 0$, then the above metric reduces to Bianchi type I spacetimes metric and a classification of Bianchi type I spacetimes according to teleparallel CKVFs is presented in [20]. Throughout this paper we will consider $q$ as a non zero constant. Using the procedure given in section 2 for the metric (3.1), the tetrad $h^a_\mu$ and its inverse $h_a^\mu$ can be obtained as:

$$h^a_\mu = \text{diag}\left(1, A(t), e^{qx}B(t), e^{qx}B(t)\right), \quad h_a^\mu = \text{diag}\left(1, \frac{1}{A(t)}, \frac{e^{-qx}}{B(t)}, \frac{e^{-qx}}{B(t)}\right) \qquad (3.2)$$

Also, the non zero components of torsion tensor turned out to be:

$$T^1_{01} = \frac{A'}{A}, \quad T^2_{02} = T^3_{03} = \frac{B'}{B}, \quad T^2_{12} = T^3_{13} = q, \qquad (3.3)$$

where a prime on a metric function denotes its derivative with respect to $t$. Using the above torsion components and the metric functions given in (3.1), Eq. (1.2) produces the following ten coupled partial differential equations, known as teleparallel conformal Killing's equations.

$$\xi^0_{,0} = \psi(t, x, y, z), \qquad (3.4)$$

$$A^2 \xi^1_{,0} - \xi^0_{,1} + AA'\xi^1 = 0, \qquad (3.5)$$

$$e^{2qx}B^2 \xi^2_{,0} - \xi^0_{,2} + BB'e^{2qx}\xi^2 = 0, \qquad (3.6)$$

$$e^{2qx}B^2\xi^3_{,0} - \xi^0_{,3} + BB'e^{2qx}\xi^3 = 0, \tag{3.7}$$

$$\xi^1_{,1} = \psi(t,x,y,z), \tag{3.8}$$

$$e^{2qx}B^2\,\xi^2_{,1} + A^2\,\xi^1_{,2} + qe^{2qx}B^2\xi^2 = 0, \tag{3.9}$$

$$e^{2qx}B^2\,\xi^3_{,1} + A^2\,\xi^1_{,3} + qe^{2qx}B^2\xi^3 = 0, \tag{3.10}$$

$$\xi^2_{,2} = \psi(t,x,y,z), \tag{3.11}$$

$$\xi^2_{,3} + \xi^3_{,2} = 0, \tag{3.12}$$

$$\xi^3_{,3} = \psi(t,x,y,z). \tag{3.13}$$

Our first attempt is to solve the system of Eqs. (3.4)-(3.13) by taking $\psi = $ const. Integrating Eqs. (3.4), (3.8), (3.11) and (3.13), we obtain the following system.

$$\xi^0 = \psi\,t + D^1(x,y,z),\quad \xi^1 = \psi\,x + D^2(t,y,z),$$
$$\xi^2 = \psi\,y + D^3(t,x,z),\quad \xi^3 = \psi\,z + D^4(t,x,y),$$

where $D^1(x,y,z)$, $D^2(t,y,z)$, $D^3(t,x,z)$ and $D^4(t,x,y)$ are functions of integration. Using this system in Eq. (3.9) and then differentiating with respect to $y$ and $x$ respectively, we have $q^2\psi = 0 \Rightarrow \psi = 0$. This indicates that LRS Bianchi type V spacetimes do not admit any proper teleparallel HVF.

Now we solve the system of Eqs. (3.4)-(3.13) simultaneously for $\psi = \psi(t,x,y,z)$ to get teleparallel CKVFs. Differentiating Eq. (3.9) with respect to $z$, Eq. (3.10) with respect to $y$ and using Eq. (3.12) in the sum of the resulting two equations, we have:

$$\xi^1_{,23} = 0 \tag{3.14}$$

Subtracting Eq. (3.8) from Eq. (3.11) and differentiating the resulting equation with respect to $y$ and $z$ respectively, we get:

$$\xi^2_{,3} = y\,F^1(t,x,z) + F^2(t,x,z), \tag{3.15}$$

where $F^1(t,x,z)$ and $F^2(t,x,z)$ are functions which arise during the process of integration. Using (3.15) in Eq. (3.12), we obtain:

$$\xi^3 = -\frac{y^2}{2} F^1(t,x,z) - y F^2(t,x,z) + F^3(t,x,z), \tag{3.16}$$

$F^3(t,x,z)$ being a function of integration. Subtracting Eq. (3.11) from Eq. (3.13) and differentiating the resulting equation with respect to $z$ and $y$ respectively, we get:

$$\begin{aligned}\xi^2 &= y\left\{\frac{z^2}{2} G^1(t,x) + z G^2(t,x)\right\} + \frac{z^2}{2} G^3(t,x) + z G^4(t,x) - \frac{y^3}{6} G^1(t,x) - \frac{y^2}{2} G^3(t,x)\\ &\quad + y G^5(t,x) + G^7(t,x),\\ \xi^3 &= -\frac{y^2}{2}\left\{z G^1(t,x) + G^2(t,x)\right\} - y\left\{z G^3(t,x) + G^4(t,x)\right\} + \frac{z^3}{6} G^1(t,x) + \frac{z^2}{2} G^2(t,x)\\ &\quad + z G^5(t,x) + G^6(t,x),\end{aligned} \tag{3.17}$$

where $G^i(t,x)$ are functions of integration, for $i=1,\ldots,7$. If we differentiate Eq. (3.9) with respect to $z$ and $y$ respectively, we obtain $G^i(t,x) = e^{-qx} H^i(t)$ for $i=1,\ldots,4$. Here $H^i(t)$ are functions of integration. With these values, the system (3.17) can be re-written as:

$$\begin{aligned}\xi^2 &= y e^{-qx}\left\{\frac{z^2}{2} H^1(t) + z H^2(t)\right\} + e^{-qx}\left\{\frac{z^2}{2} H^3(t) + z H^4(t)\right\} - e^{-qx}\left\{\frac{y^3}{6} H^1(t) + \frac{y^2}{2} H^3(t)\right\}\\ &\quad + y\, G^5(t,x) + G^7(t,x),\\ \xi^3 &= -\frac{y^2}{2} e^{-qx}\left\{z H^1(t) + H^2(t)\right\} - y e^{-qx}\left\{z H^3(t) + H^4(t)\right\} + e^{-qx}\left\{\frac{z^3}{6} H^1(t) + \frac{z^2}{2} H^2(t)\right\}\\ &\quad + z G^5(t,x) + G^6(t,x),\end{aligned} \tag{3.18}$$

We use the above system in Eqs. (3.10), (3.9) and (3.7) to obtain:

$$\begin{aligned}\xi^1 &= -\frac{B^2}{A^2} e^{2qx}\left\{\frac{z^2}{2} G_x^5(t,x) + z G_x^6(t,x) + \frac{q}{2} z^2 G^5(t,x) + q z G^6(t,x)\right\}\\ &\quad - \frac{B^2}{A^2} e^{2qx}\left\{\frac{y^2}{2} G_x^5(t,x) + y G_x^7(t,x) + \frac{q}{2} y^2 G^5(t,x) + q y G^7(t,x)\right\} + G^8(t,x),\\ \xi^0 &= -\frac{y^2}{2} e^{qx} B^2\left\{\frac{z^2}{2} H_t^1(t) + z H_t^2(t)\right\} - y e^{qx} B^2\left\{\frac{z^2}{2} H_t^3(t) + z H_t^4(t)\right\} + e^{qx} B^2\left\{\frac{z^4}{24} H_t^1(t) + \frac{z^3}{6} H_t^2(t)\right\}\\ &\quad + e^{2qx} B^2\left\{\frac{z^2}{2} G_t^5(t,x) + z G_t^6(t,x)\right\} - \frac{y^2}{2} e^{qx} B B'\left\{\frac{z^2}{2} H^1(t) + z H^2(t)\right\} - y e^{qx} B B'\left\{\frac{z^2}{2} H^3(t) + z H^4(t)\right\}\\ &\quad + e^{qx} B B'\left\{\frac{z^4}{24} H^1(t) + \frac{z^3}{6} H^2(t)\right\} + e^{qx} B B'\left\{\frac{z^2}{2} G^5(t,x) + z G^6(t,x)\right\} + F^4(t,x,y) \end{aligned} \tag{3.19}$$

The following values can be obtained if we differentiate Eq. (3.5) with respect to $y$ and $z$ respectively and then use the identity (3.14) in the resulting equation.

$$H^1(t) = \frac{c_1}{B}, \quad H^2(t) = \frac{c_2}{B}, \quad H^3(t) = \frac{c_3}{B}, \quad H^4(t) = \frac{c_4}{B}. \tag{3.20}$$

Using (3.18), (3.19) and (3.20) in Eqs. (3.6) and (3.11), we obtain the following form of teleparallel CKVFs along with conformal factor, up to some unknown functions of $t$ and $x$.

$$\xi^0 = e^{2qx} B^2 \left\{ \frac{y^2 + z^2}{2} G_t^5(t,x) + z G_t^6(t,x) + y G_t^7(t,x) \right\} +$$

$$e^{2qx} BB' \left\{ \frac{y^2 + z^2}{2} G^5(t,x) + z G^6(t,x) + y G^7(t,x) \right\} + G^9(t,x),$$

$$\xi^1 = -\frac{B^2}{A^2} e^{2qx} \left\{ \frac{z^2}{2} G_x^5(t,x) + z G_x^6(t,x) + \frac{q}{2} z^2 G^5(t,x) + q z G^6(t,x) \right\}$$

$$-\frac{B^2}{A^2} e^{2qx} \left\{ \frac{y^2}{2} G_x^5(t,x) + y G_x^7(t,x) + \frac{q}{2} y^2 G^5(t,x) + q y G^7(t,x) \right\} + G^8(t,x),$$

$$\xi^2 = \frac{y}{B} e^{-qx} \left( \frac{c_1}{2} z^2 + c_2 z \right) + \frac{1}{B} e^{-qx} \left( \frac{c_3}{2} z^2 + c_4 z \right) - \frac{1}{B} e^{-qx} \left( \frac{c_1}{6} y^3 + \frac{c_3}{2} y^2 \right) + y G^5(t,x) + G^7(t,x),$$

$$\xi^3 = -\frac{y^2}{2B} e^{-qx} (c_1 z + c_2) - \frac{y}{B} e^{-qx} (c_3 z + c_4) + \frac{1}{B} e^{-qx} \left( \frac{c_1}{6} z^3 + \frac{c_2}{2} z^2 \right) + z G^5(t,x) + G^6(t,x),$$

$$\psi(t,x,y,z) = \frac{1}{B} e^{-qx} \left( \frac{c_1}{2} z^2 + c_2 z \right) - \frac{1}{B} e^{-qx} \left( \frac{c_1}{2} y^2 + c_3 y \right) + G^5(t,x). \tag{3.21}$$

The above $\xi^a$ satisfy the Eqs. (3.6), (3.7) and (3.9)-(3.13) identically. Using these values of $\xi^a$ in Eqs. (3.4), (3.5) and (3.8), we get $c_1 = 0$ and the following integrability conditions are generated.

$$B^2 G_{tt}^5(t,x) + 3BB' G_t^5(t,x) + (BB'' + B'^2) G^5(t,x) = 0, \tag{3.22}$$

$$B^3 G_{tt}^6(t,x) + 3B^2 B' G_t^6(t,x) + B(BB'' + B'^2) G^6(t,x) = c_2 e^{-3qx}, \tag{3.23}$$

$$B^3 G_{tt}^7(t,x) + 3B^2 B' G_t^7(t,x) + B(BB'' + B'^2) G^7(t,x) = -c_3 e^{-3qx}, \tag{3.24}$$

$$G_{xx}^5(t,x) + 3q G_x^5(t,x) + 2q^2 G^5(t,x) = 0, \tag{3.25}$$

$$B^3 G_{xx}^6(t,x) + 3q B^3 G_x^6(t,x) + 2q^2 B^3 G^6(t,x) = -c_2 A^2 e^{-3qx}, \tag{3.26}$$

$$B^3 G_{xx}^7(t,x) + 3q B^3 G_x^7(t,x) + 2q^2 B^3 G^7(t,x) = c_3 A^2 e^{-3qx}, \tag{3.27}$$

$$2ABG_{tx}^{5}(t,x)+3qABG_{t}^{5}(t,x)-(A'B-3AB')G_{x}^{5}(t,x)-q(A'B-4AB')G^{5}(t,x)=0, \qquad (3.28)$$

$$2ABG_{tx}^{6}(t,x)+3qABG_{t}^{6}(t,x)-(A'B-3AB')G_{x}^{6}(t,x)-q(A'B-4AB')G^{6}(t,x)=0, \qquad (3.29)$$

$$2ABG_{tx}^{7}(t,x)+3qABG_{t}^{7}(t,x)-(A'B-3AB')G_{x}^{7}(t,x)-q(A'B-4AB')G^{7}(t,x)=0, \qquad (3.30)$$

$$A^{2}G_{t}^{8}(t,x)-G_{x}^{9}(t,x)+AA'G^{8}(t,x)=0, \qquad (3.31)$$

$$G_{t}^{9}(t,x)=G^{5}(t,x), \qquad (3.32)$$

$$G_{x}^{8}(t,x)=G^{5}(t,x), \qquad (3.33)$$

To obtain the explicit form of teleparallel CKVFs, we need to solve the above integrability conditions. It can be noticed that Eqs. (3.22)-(3.33) are highly non linear and cannot be solved directly as they stand. In next section, we present the solution of these equations in some particular cases.

4. **Particular Cases**

**Case (I):** In this case, we solve Eqs. (3.22)-(3.33) by assuming that $G^{k}(t,x)=R^{k}(t)+H^{k}(x)$, for $k=5,6,7$ and $BB''+B'^{2}=0$, so that $B(t)=\sqrt{2(c_{5}t+c_{6})}$. Using this assumption in Eq. (3.25) and differentiating it with respect to $t$, we get $G_{t}^{5}(t,x)=0 \Rightarrow G^{5}=G^{5}(x)$. Substituting back this value in Eq. (3.25), we have $G_{xx}^{5}(x)+3qG_{x}^{5}(x)+2q^{2}G^{5}(x)=0$, which is a second order differential equation and can be easily solved to get $G^{5}(x)=c_{7}e^{-qx}+c_{8}e^{-2qx}$. Using this value of $G^{5}(x)$ in Eqs. (3.32) and (3.33) and then integrating the resulting equations with respect to $t$ and $x$ respectively, we obtain $G^{9}(t,x)=c_{7}te^{-qx}+c_{8}te^{-2qx}+E^{1}(x)$ and $G^{8}(t,x)=-\frac{c_{7}}{q}e^{-qx}-\frac{c_{8}}{2q}e^{-2qx}+E^{2}(t)$, where $E^{1}(x)$ and $E^{2}(t)$ are functions of integration. Substituting these values in Eq. (3.31) and differentiating it with respect to $t$ and $x$ repeatedly, we obtain $c_{8}=0$, $A(t)=\sqrt{q^{2}t^{2}+2c_{9}t+c_{10}}$, $E^{1}(x)=\frac{c_{7}c_{9}}{q^{2}}e^{-qx}+c_{12}$ and $E^{2}(t)=\frac{c_{13}}{A}$. Hence the values of $G^{5}$, $G^{8}$ and $G^{9}$ become $G^{5}(x)=c_{7}e^{-qx}$, $G^{9}(t,x)=c_{7}te^{-qx}+\frac{c_{7}c_{9}}{q^{2}}e^{-qx}+c_{12}$ and $G^{8}(t,x)=-\frac{c_{7}}{q}e^{-qx}+\frac{c_{13}}{A}$. Simplifying Eq. (3.28) by using the value of $G^{5}$, we get $c_{5}=0$. Thus the metric function $B$ becomes constant. Moreover, if we differentiate Eqs. (3.23) and (3.24) with respect to $x$, we find that $c_2 = c_3 = 0$. Putting $c_2 = 0$ in Eq. (3.26) and differentiating it with respect to $t$, we get $G_t^6(t,x) = 0 \Rightarrow G^6 = G^6(x)$. Using this values in Eq. (3.26), we have $G_{xx}^6(x) + 3q\, G_x^6(x) + 2q^2\, G^6(x) = 0$, which implies $G^6(x) = c_{14} e^{-qx} + c_{15} e^{-2qx}$. Putting this value of $G^6$ in Eq. (3.29), we obtain $c_{15} = 0$. Hence we have $G^6(x) = c_{14} e^{-qx}$. Similarly, the simultaneous solution of Eqs. (3.24), (3.27) and (3.30) gives $G^7(x) = c_{16} e^{-qx}$. Hence all the unknown functions involved in the integrability conditions (3.22)-(3.33) have been explicitly found, which are given below:

$$G^5(x) = c_7 e^{-qx}, \quad G^6(x) = c_{14} e^{-qx}, \quad G^7(x) = c_{16} e^{-qx},$$
$$G^9(t,x) = c_7 t e^{-qx} + \frac{c_7 c_9}{q^2} e^{-qx} + c_{12}, \quad G^8(t,x) = -\frac{c_7}{q} e^{-qx} + \frac{c_{13}}{A}. \tag{4.1}$$

With these values, the system (3.21) becomes:

$$\xi^0 = c_7 t e^{-qx} + \frac{c_7 c_9}{q^2} e^{-qx} + c_{12},$$
$$\xi^1 = -\frac{c_7}{q} e^{-qx} + \frac{c_{13}}{A},$$
$$\xi^2 = e^{-qx}(c_4 z + c_7 y + c_{16}), \tag{4.2}$$
$$\xi^3 = e^{-qx}(-c_4 y + c_7 z + c_{14}),$$
$$\psi(t,x,y,z) = c_7 e^{-qx},$$

where the metric functions $A$ and $B$ get the values $A(t) = \sqrt{q^2 t^2 + 2c_9 t + c_{10}}$ and $B = const.$ From above we can see that for these values of the metric functions, the LRS Bianchi type V spacetimes admit six teleparallel CKVFs, out of which one is a proper teleparallel CKVF which can be expressed as ( by choosing $c_9 = 1$ ), $\xi_1 = \left(t + \frac{1}{q^2}\right) e^{-qx} \partial_t - \frac{1}{q} e^{-qx} \partial_x + y e^{-qx} \partial_y + z e^{-qx} \partial_z$.

The remaining five are teleparallel KVFs with the following generators:

$$\xi_2 = \partial_t, \quad \xi_3 = \frac{1}{A}\partial_x, \quad \xi_4 = z e^{-qx}\partial_y - y e^{-qx}\partial_z, \quad \xi_5 = e^{-qx}\partial_y, \quad \xi_6 = e^{-qx}\partial_z. \tag{4.3}$$

It is worth noticing that in general relativity, LRS Bianchi type V spacetimes admit six CKVFs in which four are the KVFs and one is each a proper HVF and a proper CKVF [19], for $A = A(t)$ and $B = const$. But in the context of teleparallel gravity, the same spacetimes admit six teleparallel CKVFs with one proper teleparallel CKVF and five teleparallel KVFs, while no proper HVF exists. This increase of one KVF and the disappearance of HVF is because of the presence of non zero torsion and vanishing curvature in the spacetimes.

**Case (II):** In this case we solve the integrability conditions (3.22)-(3.33) by taking $B = const.$ and $A = A(t)$ such that it satisfies the differential constraint $AA'' - A'^2 = 0$, which implies that $A(t) = c_5 e^{c_6 t}$ with $c_5 \neq 0$ and $c_6 \neq 0$. For simplicity, we choose $c_5 = c_6 = 1$. With these values, Eq. (3.22) gives $G^5(t,x) = t E^1(x) + E^2(x)$, $E^1(x)$ and $E^2(x)$ being functions of integration. Putting this value of $G^5(t,x)$ in Eq. (3.28) and differentiating the resulting equation with respect to $t$, we have $G^5(t,x) = (c_7 t + c_7 q x + c_8) e^{-qx}$. If we put this value of $G^5(t,x)$ in Eq. (3.25), it gives $c_7 = 0$ and hence $G^5(t,x) = c_8 e^{-qx}$. Integrating Eqs. (3.32) and (3.33), one can easily find that $G^8(t,x) = -\frac{c_8}{q} e^{-qx} + E^3(t)$ and $G^9(t,x) = c_8 t e^{-qx} + E^4(x)$, where $E^3(t)$ and $E^4(x)$ are functions of integration. The functions $G^5$, $G^8$ and $G^9$ can be re-written in the following form if we differentiate Eq. (3.31) with respect to $x$ and $t$ respectively.

$$G^5(t,x) = 0, \quad G^8(t,x) = -c_9 e^{-2t} + c_{14} e^{-t}, \quad G^9(t,x) = c_9 x + c_{10}. \qquad (4.4)$$

Integrating Eq. (3.23) twice with respect to $t$, we get: $G^6(t,x) = \frac{c_2}{2} t^2 e^{-3qx} + t E^5(x) + E^6(x)$, $E^5(x)$ and $E^6(x)$ being functions of integration. Putting this value of $G^6(t,x)$ in Eq. (3.29) and differentiating the resulting equation twice with respect to $t$, we have $c_2 = 0$ and $G^6(t,x) = (c_{11} t + c_{11} q x + c_{12}) e^{-qx}$. Using this value in Eq. (3.26), we obtain $c_{11} = 0$ and hence $G^6(t,x) = c_{12} e^{-qx}$. Following the same procedure for the solution of Eqs. (3.24), (3.27) and (3.30), we have $c_3 = 0$ and $G^7(t,x) = c_{13} e^{-qx}$. Putting back all these values in (3.21), we have:

$$\xi^0 = c_9\,x + c_{10},$$
$$\xi^1 = -c_9\,e^{-2t} + c_{14}\,e^{-t},$$
$$\xi^2 = (c_4 z + c_{13})\,e^{-qx}, \tag{4.5}$$
$$\xi^3 = (-c_4 y + c_{12})\,e^{-qx},$$
$$\psi(t, x, y, z) = 0.$$

Since the conformal factor vanishes, it shows that in this case the teleparallel CKVFs are just teleparallel KVFs. The number of teleparallel KVFs is clearly six with no proper teleparallel HVF. The generators of the above six teleparallel KVFs can be written as:

$$\xi_1 = x\partial_t - e^{-2t}\partial_x, \quad \xi_2 = \partial_t, \quad \xi_3 = e^{-t}\partial_x,$$
$$\xi_4 = ze^{-qx}\partial_y - ye^{-qx}\partial_z, \quad \xi_5 = e^{-qx}\partial_y, \quad \xi_6 = e^{-qx}\partial_z \tag{4.6}$$

**Case (III):** Here we solve the integrability conditions (3.22)-(3.33) by taking $A = \text{const.}$ and $B = B(t)$ such that $BB'' + B'^2 = 0$, which implies that $B(t) = \sqrt{2(c_5 t + c_6)}$ with $c_5 \neq 0$. Solving the system of equation (3.22)-(3.33) for these values of the metric functions we see that the conformal factor $\psi$ again vanishes, showing that there is no proper teleparallel CKVF and the teleparallel CKVFs in this case again reduce to teleparallel KVFs, which are given below.

$$\xi^0 = c_7\,x + c_8,$$
$$\xi^1 = c_7\,t + c_9,$$
$$\xi^2 = \frac{e^{-qx}}{\sqrt{(c_5 t + c_6)}}\left(\frac{c_4 z}{\sqrt{2}} - \frac{2c_{10}}{c_5}\right), \tag{4.7}$$
$$\xi^3 = -\frac{e^{-qx}}{\sqrt{(c_5 t + c_6)}}\left(\frac{c_4 y}{\sqrt{2}} + \frac{2c_{11}}{c_5}\right).$$

choosing $c_5 = 1$, the above six teleparallel KVFs have the following generators:

$$\xi_1 = x\partial_t + t\partial_x, \quad \xi_2 = \partial_t, \quad \xi_3 = \partial_x, \quad \xi_4 = \frac{ze^{-qx}}{\sqrt{2(t+c_6)}}\partial_y - \frac{ye^{-qx}}{\sqrt{2(t+c_6)}}\partial_z,$$

$$\xi_5 = -\frac{2e^{-qx}}{\sqrt{(t+c_6)}}\partial_y, \quad \xi_6 = -\frac{2e^{-qx}}{\sqrt{(t+c_6)}}\partial_z \tag{4.8}$$

Note that for $A = \text{const.}$ and $B = B(t)$, the same spacetimes admit six HVFs in the context of general relativity, out of which five are KVFs and one is a proper HVF [19]. The disappearance of proper HVF and the increase of one KVF in teleparallel gravity shows the effect of non zero torsion and vanishing curvature on symmetries of the LRS Bianchi type V spacetimes.

**Case (IV):** In this case we take $A(t) = B(t)$ such that $AA'' + A'^2 = 0$, that is $A(t) = \sqrt{2(c_5 t + c_6)}$ with $c_5 \neq 0$. Under these restrictions, the simultaneous solution of the system of equations (3.22)-(3.33) gives $\psi = 0$. This shows that the teleparallel CKVFs in this case are just KVFs, which are listed below.

$$\xi^0 = c_7 x + c_8,$$
$$\xi^1 = -\frac{1}{\sqrt{c_5 t + c_6}}\left\{\frac{2c_{10} q z}{c_5} + \frac{2c_{11} q y}{c_5} - \frac{c_9}{\sqrt{2}}\right\} + \frac{c_7}{c_5},$$
$$\xi^2 = \frac{1}{\sqrt{c_5 t + c_6}}\left\{\frac{c_4 z e^{-qx}}{\sqrt{2}} - \frac{2(c_{12} e^{-qx} + c_{11} e^{-2qx})}{c_5}\right\}, \quad (4.9)$$
$$\xi^3 = -\frac{1}{\sqrt{c_5 t + c_6}}\left\{\frac{c_4 y e^{-qx}}{\sqrt{2}} + \frac{2(c_{13} e^{-qx} + c_{10} e^{-2qx})}{c_5}\right\}.$$

In this case LRS Bianchi type V spacetimes admit eight teleparallel KVFs, whose generators can be written as ( by choosing $c_5 = 1$):

$$\xi_1 = x\partial_t + \partial_x, \quad \xi_2 = \partial_t, \quad \xi_3 = \frac{1}{\sqrt{2(t+c_6)}}\partial_x,$$
$$\xi_4 = \frac{z e^{-qx}}{\sqrt{2(t+c_6)}}\partial_y - \frac{y e^{-qx}}{\sqrt{2(t+c_6)}}\partial_z,$$
$$\xi_5 = -\frac{2 e^{-qx}}{\sqrt{(t+c_6)}}\partial_z, \quad \xi_6 = -\frac{2qz}{\sqrt{(t+c_6)}}\partial_x - \frac{2 e^{-2qx}}{\sqrt{(t+c_6)}}\partial_z, \quad (4.10)$$
$$\xi_7 = -\frac{2qy}{\sqrt{(t+c_6)}}\partial_x - \frac{2 e^{-2qx}}{\sqrt{(t+c_6)}}\partial_y, \quad \xi_8 = -\frac{2 e^{-qx}}{\sqrt{(t+c_6)}}\partial_y.$$

**Case (V):** Finally we consider $A = \text{const.}$ and $B = \text{const.}$. It is straightforward to Solve Eqs. (3.22)-(3.33) for these values of the metric function and obtain the following system:

$$\xi^0 = c_5 x + c_6,$$
$$\xi^1 = c_7 q z + c_8 q y + c_5 t + c_9,$$
$$\xi^2 = c_4 z e^{-qx} + c_{10} e^{-qx} + c_8 e^{-2qx}, \quad (4.11)$$
$$\xi^3 = -c_4 y e^{-qx} + c_{11} e^{-qx} + c_7 e^{-2qx},$$
$$\psi = 0.$$

Like the previous three cases, again the conformal factor $\psi$ vanishes and the teleparallel CKVFs are reduced to teleparallel KVFs. The number of teleparallel KVFs turned out to be eight with the following generators:

$$\xi_1 = z e^{-qx} \partial_y - y e^{-qx} \partial_z, \; \xi_2 = x\partial_t + t\partial_x, \; \xi_3 = \partial_t, \; \xi_4 = qz\partial_x + e^{-2qx}\partial_z,$$

$$\xi_5 = qy\partial_x + e^{-2qx}\partial_y, \; \xi_6 = \partial_x, \; \xi_7 = e^{-qx}\partial_y, \; \xi_8 = e^{-qx}\partial_z. \qquad (4.12)$$

## 5. **Summary**

In this paper we have investigated teleparallel homothetic and conformal Killing vector fields for LRS Bianchi type V spacetimes in the presence of non zero torsion and vanishing curvature. In some cases, the obtained results are compared with those of general relativity. The first outcome of our study is that the LRS Bianchi type V spacetimes do not admit any proper teleparallel HVF. The general form of teleparallel CKVFs is found, subject to some integrability conditions. Due to the high non-linearity of integrability conditions, we were unable to solve them generally. However, in some particular cases these integrability conditions are completely solved to get the final form of teleparallel CKVFs. In the first case, the solution of integrability conditions produces a proper teleparallel CKVF along with five teleparallel KVFs, while in the remaining four cases the teleparallel CKVFs are reduced to teleparallel KVFs. The number of teleparallel KVFs for LRS Bianchi type V spacetimes turned out to be six or eight.

## **References**

[1]   H. Weyl, Gravitation und Elektrizitat, Sitzungsber. Akad. Wiss. Berlin, p. 465 (1918).

[2]   A. Einstein, Auf die Riemann-Metrik und den Fern-Parallelismus gegrundete einheitliche Feldtheorie, *Math. Annal.* **102**, 685 (1930).

[3]   C. Moller, K. Dan. *Vidensk. Selsk. Mat. Fys. Skr.* **1**, 10 (1961).

[4]   C. Pellegrini and J. Plebanski, K. Dan. *Vidensk. Selsk. Mat. Fys. Skr.* **2**, 4 (1962).

[5]   K. Hayashi and T. Nakano, *Prog. Theor. Phys.* **38**, 491 (1967).

[6]   K. Hayashi, *Phys. Lett. B* **69**, 441 (1977).

[7]   K. Hayashi and T. Shirafuji, *Phys. Rev. D* **19**, 3524 (1979).

[8]   R. Weitzenböck, Invarianten Theorie (Noordhoft, 1923).

[9]   G. S. Hall, Symmetries and curvature structure in general relativity, World Scientific, 2004.

[10]  A. Z. Petrov, Einstein spaces, Oxford University Press, 1969.

[11]  M. Sharif and M. J. Amir, *Mod. Phys. Lett. A* **23**, 963 (2008)

[12]  R. Maartens and S. D. Maharaj, *Class. Quantum Grav.* **3**, 1005 (1986)

[13]  R. Maartens, S. D. Maharaj and B. O. J. Tupper, *Class. Quantum Grav.* **12**, 2577 (1995)

[14]  S. Moopanar and S. D. Maharaj, *Inter. J. Theor. Phys.* **49**, 1878 (2010)

[15]  K. Saifullah and S. Yazdan, *Int. J. Mod. Phys. D* **18**, 71 (2009)

[16]  R. Maartens and S. D. Maharaj, *Class. Quantum Grav.* **8**, 503 (1991)

[17]  D. Kramer and J. Carot, *J. Math. Phys.* **32**, 1857 (1991)

[18]  G. S. Hall and J. Carot, *Class. Quantum Grav.* **11**, 475 (1994)

[19]  S. Khan, T. Hussain, A. H. Bokhari and G. A. Khan, *Commun. Theor. Phys.* **65**, 315 (2016)

[20]  G. Shabbir and H. Khan, *Rom. J. Phys.* **59**, 79 (2014)

[21]  G. Shabbir, A. Khan and S. Khan, *Int. J. Theor. Phys.* **52**, 1182 (2013)

[22]  S. Khan, T. Hussain and G. A. Khan, *Eur. Phys. J. Plus* **129**, 228 (2014)

[23]  S. Khan, T. Hussain and G. A. Khan, *Int. J. Geom. Meth. Mod. Phys.* (2015) DOI:10.1142/S0219887816500304

[24]  R. P. Singh and L. Ladav, *Rom. Rep. Phys.*, **63** 587 (2011)

[25]  H. Stephani, D. Kramer, M. Maccallum, C. Hoenselaers and E. Herlt, Exact Solutions of Einstein's Field Equations, Second Edition, Cambridge University Press, 2003.